\documentclass[twoside]{article}

\usepackage{float}
\usepackage{color}
\usepackage{greekbf}
\usepackage{float}
\usepackage{amsmath,amssymb}
\usepackage{mathrsfs}
\usepackage{enumerate}
\usepackage{srcltx}
\usepackage{mathtools}
    \mathtoolsset{showonlyrefs}
\usepackage{graphicx, psfrag}
\usepackage{hyperref}
\hypersetup{bookmarks=false,
colorlinks=false,
linkcolor=black,
urlcolor=black,
pagebackref=true,
pdfstartview={FitH},
bookmarksopenlevel=2
}

\DeclareMathAlphabet{\mathbf}{OT1}{cmr}{bx}{it}

\definecolor{red}{rgb}{0.9,0,0}
\definecolor{blue}{rgb}{0.2,0.2,0.8}
\definecolor{green}{rgb}{0.0,0.5,0.2}
\definecolor{darkblue}{rgb}{0.2,0.2,0.5}
\definecolor{orange}{rgb}{1,0.5,0}

\newcommand {\xib}{\mathbf{\xi}}

\newcommand {\Tc}  {\mathcal{T}}

\begin{document}

\title{\vspace{-3cm} {\bf An analytical benchmark for MD codes:  \\ testing LAMMPS on the 2nd generation Brenner potential}}

\author{
Antonino Favata$^1$\!\!\!\!\! \and Andrea Micheletti$^2$\!\!\!\!\! \and Seunghwa Ryu$^{3}$\!\!\!\!\! \and  Nicola M. Pugno$^{4,5,6}$
}

\date{\today}

\maketitle

\vspace{-1cm}
\begin{center}
{\small
$^1$ Department od Structural Engineering and Geotechnics\\ Sapienza University of Rome, Italy\\[2pt]
\href{mailto:antonino.favata@uniroma1.it}{antonino.favata@uniroma1.it}\\[8pt]
$^2$ Dipartimento di Ingegneria Civile e Ingegneria Informatica\\
University of Rome TorVergata, Italy\\[2pt]
\href{mailto:micheletti@ing.uniroma2.it}{micheletti@ing.uniroma2.it}\\[8pt]
$^3$ Korea Advanced Institute of Science and Technology (KAIST)\\[2pt]
\href{mailto:ryush@kaist.ac.kr}{ryush@kaist.ac.kr}\\[8pt]
$^4$ Laboratory of Bioinspired and Graphene Nanomechanics\\ Department of Civil, Environmental and Mechanical Engineering\\ University of Trento, Italy\\[2pt]
\href{mailto:nicola.pugno@unitn.it}{nicola.pugno@unitn.it}\\[8pt]
$^5$ Center for Materials and Microsystems\\
Fondazione Bruno Kessler,  Trento, Italy\\[8pt]
$^6$ School of Engineering and Materials Science\\
Queen Mary University of London,  UK
}

\end{center}

\pagestyle{myheadings}
\markboth{A.~Favata, A.~Micheletti, S.~Ryu, N.M.~Pugno}
{MD Benchmark: Testing LAMMPSs}

\vspace{-0.5cm}
\section*{Abstract}
An analytical benchmark is proposed for  graphene and carbon nanotubes, that may serve to test whatsoever  molecular dynamics code implemented with REBO potentials. By exploiting the benchmark,  we checked results produced by LAMMPS (Large-scale Atomic/Molecular Massively Parallel Simulator)  when adopting the second generation Brenner potential, we made evident that the code in its current implementation produces  results which are offset from those of the benchmark by a significant amount, and provide evidence of the reason.

\vspace{0.7cm}
\noindent \textbf{Keywords}: REBO potentials, 2nd generation Brenner potential, LAMMPS, Benchmark

\vspace{0.7cm}

\section{Introduction}
Molecular dynamics (MD) simulations are nowadays more and more popular in scientific applications, especially in those fields of material science involving nanotechnology and advanced material design.  On one side, there are advantages in the speed and accuracy of the simulations, with the model of the potential for atomic interactions being optimized to reproduce either experimental values or quantities extimated by first principles calculations (considered, as a matter of facts, just like experimental results). On the other side, it is more and more frequent to use commercial or open-access codes implementing off-the-shelf potential models, and use them as a black box, without having a precise feeling with the code itself. One of the most used simulator is LAMMPS (Large-scale Atomic/Molecular Massively Parallel Simulator), able to implement several interatomic potentials.
By using an analytical discrete mechanical model, we present a benchmark for the equilibrium problem of graphene and carbon nanotubes, which can be applied to any kind of REBO (reactive empirical bond-order) potential. The analytical condition proposed produces results in complete agreement with First Principles, Density Functional Theory and Monte Carlo simulations.
With the aid of this benchmark, we show that LAMMPS code, when implemented with the second generation Brenner potential,  produces results which are offset from those of the benchmark by a significant amount, and provide evidence of the reason.
The purpose of this letter it is not to just  to supply a test for the LAMMPS code, rather, it is to provide a general tool for testing any MD code.

\section{An analytical discrete model for equilibrium configurations of FGS\lowercase{s} and CNT\lowercase{s}}\label{model}
The benchmark solution we propose has been developed within the context of carbon macromolecules, such as Flat Graphene Strips (FGSs) or Carbon Nanotubes (CNTs). When regarded from the point of view of MD, such aggregates are  modelled as  sets of mass points, whose configuration is described by the Cartesian coordinates of each point with respect to a chosen reference frame; each point is then interacting with the others -- at least with the closest ones -- and the interaction is captured by a suitable empirical potential, whose shape and parameters are fitted with a set of selected experiments and \textit{ab initio} calculations. The last generation potentials usually take into account multi-particle interactions, up to the third nearest neighbor, which is indispensable to capture the mechanics of complex systems, such as carbon macromolecules.

In order to provide an easy-to-visualize mechanical picture, the perspective we here adopt is not the one of MD, 
we consider instead the approach of Favata \textit{et al.}\cite{Favata_2014a}, where a discrete mechanical model is detailed for 2D carbon allotropes. In this view,  the configuration of a molecular aggregate is not identified by the coordinates of the mass points, but rather by a suitable finite list of order parameters. In particular, the conditions of \textit{natural equilibrium} of the aggregate can be determined and expressed in terms of such list and independently of  the choice of the REBO potential. As we will see, the prediction of such equations are in total agreement with First Principles, Density Functional Theory and Monte Carlo simulations; moreover, given their generality, they can be exploited to establish benchmark solutions. 

\begin{figure}[h!]
	\centering
	\includegraphics[width=0.46\textwidth]{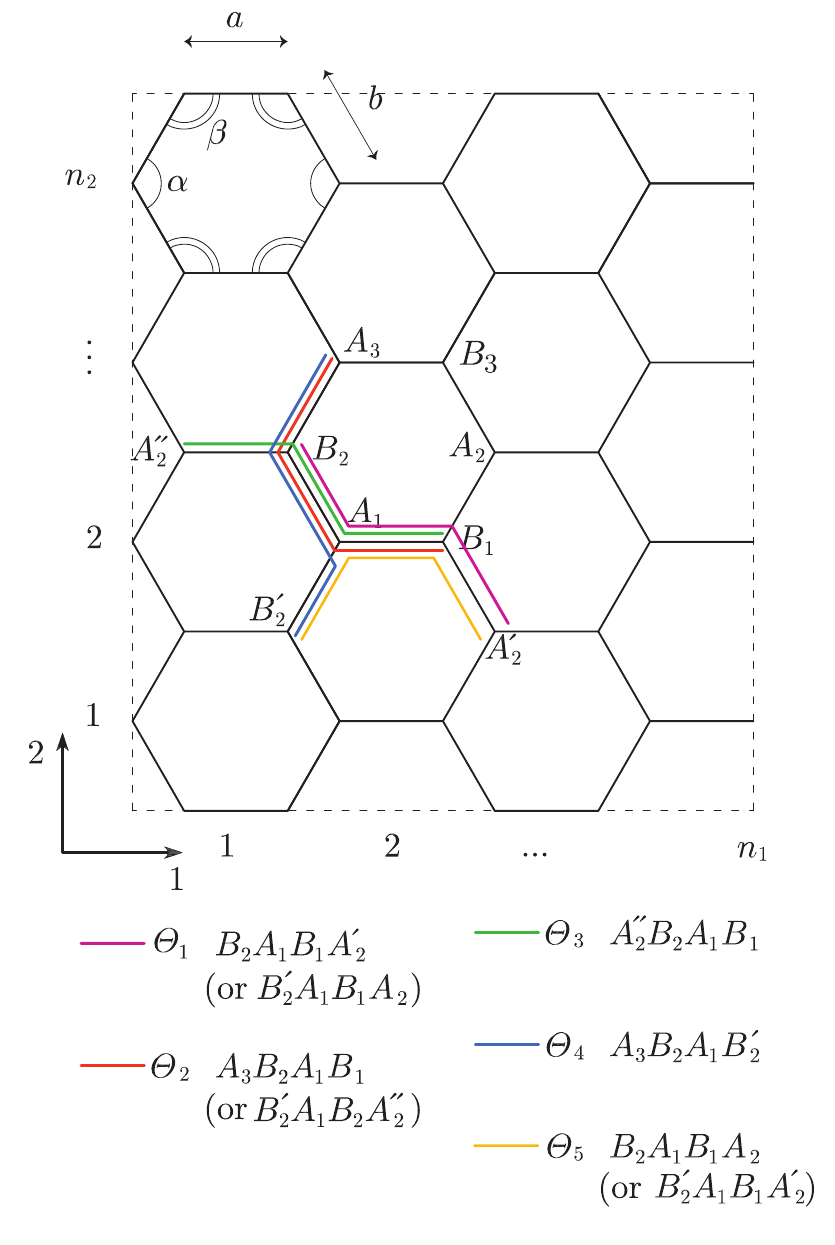}
	\caption{Order parameters in a graphene sheet.}
	\label{rectangle}
\end{figure}

In order to understand the physical meaning of the conditions we propose, we 
summarize some of the results of Favata \textit{et al.}\cite{Favata_2014a}. Starting with the geometry, with reference to Fig. \ref{rectangle}, 
let the axes 1 and 2 be respectively aligned with the armchair and zigzag directions, and let $n_1, n_2$ be the number of hexagonal cells counted along these axes.
Let us consider now the representative cell $A_1 B_1 A_2 B_3 A_3 B_2 A_1$. We note that the sides $\overline{A_1 B_1}$ and $\overline{A_3 B_3}$ are aligned with the axis 
1; the common length of corresponding bonds will be denoted by $a$, and we will call  {\em a-type} the corresponding bonds.  We see that the other four sides have equal length $b$ ({\em b-type bonds}). We pass to introduce the \textit{bond angles} and, since we intend to consider interactions up to the third neighbor,  the \textit{dihedral angles}. As to the bond angles, we notice that they can be of \emph{$\alpha$-type}  and \emph{$\beta$-type}  (e.g., respectively, $\widehat{A_3  B_2 A_1}$ and $\widehat{B_2 A_1 B_1}$; see Fig. \ref{rectangle}).  As to the dihedral angles, there are five types $(\Theta_1,\ldots,\Theta_5)$, which can be identified with the help  of the colored bond chains in Fig. \ref{rectangle}. In conclusion, to determine the deformed configuration of a representative hexagonal cell, no matter if that cell belongs to a FGS or to an achiral CNT, we need to determine  the 9-entry \emph{order-parameter substring}:
\begin{equation}\label{subs}
\boldsymbol\xi_{sub}:=(a,b,\alpha,\beta,\Theta_1,\ldots,\Theta_5)\,.
\end{equation}
The complete order-parameter string for the whole molecular aggregate can be obtained by  sequential juxtaposition of  substrings.

\vspace{1cm}

Due to the geometric compatibility conditions induced by the built-in symmetry (see Favata \textit{et al.}\cite{Favata_2014a} for details), only three  of the nine kinematic variables determine the natural configuration, which are chosen to be $a, b,$ and $\alpha$. In particular, by distinguishing the armchair (superscript A) from the zigzag (superscript Z) case, the order-parameter substring are given by, respectively:
\begin{equation}\label{subsAZ}
\begin{aligned}
\boldsymbol\xi_{sub}^A
=& ( a,b,\alpha,\widetilde{\beta}^{A}(\alpha,\varphi^A),\\
& \widetilde\Theta_1^A(\alpha,\varphi^A),\widetilde\Theta_2^A(\alpha,\varphi^A), 2\widetilde\Theta_2^A(\alpha,\varphi^A),0,0);\\
\boldsymbol\xib_{sub}^Z
=& (a,b,\alpha,\widetilde{\beta}^{Z}(\alpha,\varphi^Z),\\
&\varphi^Z, \widetilde\Theta_2^Z(\alpha,\varphi^Z),0,2\,\widetilde\Theta_2^Z(\alpha,\varphi^Z),0).
\end{aligned}
\end{equation}
The explicit form of the functions $\widetilde{\beta}^{A,Z}, \widetilde\Theta_1^A, \widetilde\Theta_2^{A,Z}$ is given in Favata \textit{et al.}\cite{Favata_2014a}
In \eqref{subsAZ},  $\varphi^{A}={\pi}/{n_1}$ is the angle between the plane of $A_1 B_1 B_3$ and the plane of $B_1 A_2 B_3$ when an armchair CNT is considered, and  $\varphi^{Z}=\frac{\pi}{n_2}$ the angle between the planes of $A_1 B_1 A_2$ and $A_2 B_3 A_3$, when a zigzag CNT up is considered. In case of a FGS, we have $\varphi^{A,Z}=0$, $\widetilde{\beta}^{A,Z}=\pi-\alpha/2$, and $\widetilde\Theta_1^A=\widetilde\Theta_2^{A,Z}\equiv 0$.

The equilibrium equations turn out to be the following ones:

\begin{equation}\label{finali}
\begin{aligned}
&\sigma_a= 0\,,\quad
\sigma_b= 0\,,\\
&\tau_\alpha+2\beta,_\alpha\, \tau_\beta +\Theta_1,_\alpha\,\Tc_1+2\Theta_2,_\alpha\,\Tc_2+\Theta_3,_\alpha\,\Tc_3+\Theta_4,_\alpha\,\Tc_4\,,
\end{aligned}
\end{equation}
where  $\sigma_a, \sigma_b, \tau_\alpha, \tau_\beta$, and $\Tc_i$,  are the so-called {\em nanostresses},  work-conjugate to changes of, respectively, bond lengths, bond angles, and dihedral angles of each type considered.
The form of the third of \eqref{finali} depends on which of the two achiral CNTs is dealt with; more precisely, we have that
\begin{equation}\label{mairichiamata}
\begin{aligned}
\tau_\alpha^{A}+2\,\beta^{A},_\alpha\, \tau_\beta^{A} +\Theta_1^{A},_\alpha\,\Tc_1^{A}+2\Theta_2^{A},_\alpha\,\Tc_2^{A}+\Theta_3^{A},_\alpha\,\Tc_3^{A} &= 0\,,\\
\tau_\alpha^{Z}+2\,\beta^{Z},_\alpha\, \tau_\beta^{Z} +2{\Theta_2^{Z}},_\alpha\Tc_2^{Z}+{\Theta_4^{Z}},_\alpha\Tc_4^{Z}&= 0\,.
\end{aligned}
\end{equation}
Due to their generality, the  conditions \eqref{finali} may serve as a benchmark for any REBO potential. To express the equilibrium equations in terms of the Lagrangian coordinates  $a, b,$ and $\alpha$, it is necessary to introduce the constitutive equations for the stress, which result from the assignment of an intermolecular potential.
In the next section,  we detail the formulas in the Brenner 2nd generation REBO potential \cite{Brenner_2002} which are needed to solve \eqref{finali} in terms of the order parameters.

\section{REBO potentials}\label{potentials}

In the Brenner 2nd generation REBO potential, the  binding energy $V^{\rm REBO}$ of a molecular aggregate is written as a sum over nearest neighbors:
\begin{equation}\label{V}
V^{\rm REBO}=\sum_i \sum_{J<I} V_{IJ}\,;
\end{equation}
the interatomic potential $V_{IJ}$ is given by the construct
\begin{equation}\label{Vij}
V_{IJ}=V_R(r_{IJ})+b_{IJ}V_A(r_{IJ})\,,
\end{equation}
where the individual effects of the \emph{repulsion} and \emph{attraction functions} $V_R(r_{IJ})$ and $V_A(r_{IJ})$, which model pair-wise interactions of  atoms $I$ and $J$ depending on their distance $r_{IJ}$, are modulated by the \emph{bond-order function} $b_{IJ}$. The repulsion and attraction functions have the following form:
\begin{equation}\label{VA}
\begin{aligned}
V_A(r)&=f^C(r)\sum_{n=1}^{3}B_n e^{-\delta_n r}\,,\\
V_R(p)&=f^C(r)\left( 1 + \frac{Q}{r} \right) A e^{-\gamma r}\,,
\end{aligned}
\end{equation}
where $f^C(r)$ is a \emph{cutoff function} limiting the range of covalent interactions to nearest neighbors, and where $Q$, $A$, $B_n$, $\gamma$, and $\delta_n$, are parameters to be chosen fit to a material-specific dataset. The remaining ingredient in \eqref{Vij} is the \emph{bond-order function}:
\begin{equation}\label{bij}
b_{IJ}=\frac{1}{2}(b_{IJ}^{\sigma-\pi}+b_{JI}^{\sigma-\pi})+b_{IJ}^\pi\,,
\end{equation}
where superscripts $\sigma$ and $\pi$ refer to two types of bonds: the strong covalent $\sigma$-bonds, between atoms in the same plane, and the $\pi$-bonds 
, which are perpendicular to the plane of $\sigma$-bonds. We now describe the functions $b_{IJ}^{\sigma-\pi}$ and $b_{IJ}^\pi$.

The role of function $b_{IJ}^{\sigma-\pi}$ is to account for the local coordination of, and the bond angles relative to, atoms $I$ and $J$, respectively; its form is:
\begin{equation}\label{G}
\begin{aligned}
b_{IJ}^{\sigma-\pi}= &\Bigg(1+\sum_{K\neq I,J} f_{ik}^C(r_{IK})G(\cos\theta_{IJK})  \,e^{\lambda_{IJK}}+\\
& \hspace{1cm} + P_{IJ}(N_I^C,N_I^H)  \Bigg)^{-1/2}\,.
\end{aligned}
\end{equation}
Here, for each fixed pair of indices $(I,J)$, (a) the cutoff function $f_{IK}^C(r)$ limits the interactions of atom $I$ to those with its nearest neighbors; (b) $\lambda^{IJK}$ is a string of parameters designed to prevent attraction in some specific situations; (c) function $P_{IJ}$ depends on $N_I^C$ and $N_I^H$, the numbers of $C$ and $H$ atoms that are nearest neighbors of atom $I$, and is meant to adjust the bond-order function 
according to the specific environment of the C atoms in one or another molecule; (d) according to Brenner et al.\cite{Brenner_2002}, for solid-state carbon, the values of  both the string $\lambda^{IJK}$ and the function $P_{IJ}$ are taken null. Finally,   function $G(\cos\theta_{IJK})$ modulates the contribution of each nearest neighbor in terms of the cosine of the angle between the $IJ$ and $IK$ bonds; its analytic form is given by three six-order polynomial splines in $\cos\theta$, each of them defined in an interval of the bond angle $\theta$. The corresponding coefficients are determined by fitting each polynomial spline to the values of $G(\cos\theta)$ at certain values of $\theta$.

Function $b_{IJ}^\pi$ is given a split representation:
\begin{equation}
b_{IJ}^\pi=\Pi_{IJ}^{RC}+b_{IJ}^{DH};
\end{equation}
the first addendum depends on whether the bond between atoms $I$ and $J$ has a radical character and on whether it is part of a conjugated system, the second depends on dihedral angles.
Function $b_{IJ}^{DH}$ is given by
\begin{equation}
\begin{aligned}
& b_{IJ}^{DH}= T_{IJ}(N_I^t,N_J^t,N_{IJ}^{\rm conj})\times\\
&\times\left(\sum_{K(\neq I,J)}\sum_{L(\neq I,J)}\big( 1-\cos^2\Theta_{IJKL} \big)f_{IK}^C(r_{IK})f_{JL}^C(r_{JL})  \right)\,,
\end{aligned}
\end{equation}
where function $T_{IJ}$ is a tricubic spline depending on $N_I^t=N_I^C+N_I^H$, $N_J^t$, and $N_{IJ}^{\rm conj}$, a function  of local conjugation, while $\Theta_{IJKL}$ is the diedral angle between the planes of $I, J, K$ and $I, J, L$. 

When the point of view described in Sect. \ref{model} is assumed, the expressions of the potentials have to be specialized and written in terms of the order parameters in the substrings \eqref{subs}. On introducing the potentials $V_a$ and $V_b$ for the \textit{a-} and \textit{b}-type bonds, we have, respectively:

\begin{equation}\label{Va:AZ}
\begin{aligned}
V_a(a,\beta,\Theta_1)&=V_R(a)+b_a(\beta, \Theta_1)\,V_A(a)\,,\\
V_b(b,\alpha,\beta,\Theta_2,\Theta_3,\Theta_4)&=V_R(b) +
b_b(\alpha, \beta, \Theta_2, \Theta_3, \Theta_4)\,V_A(b)\,
\end{aligned}
\end{equation}
(see Favata \textit{et al.}\cite{Favata_2014a} for details).

Once this has been done, the nanostresses entering the balance equations \eqref{finali} can be expressed in terms of the order parameters by means of the following constitutive relations:
\begin{equation}\label{nanostress}
\begin{aligned}
\sigma_a & = V_R'(a)+b_a(\beta,\Theta_1)\,V_A'(a)\,,\\
\sigma_b &  = V_R'(b)+b_b(\alpha,\beta,\Theta_2,\Theta_3,\Theta_4)\,V_A'(b)\,,\\
\tau_\alpha & = b_b,_\alpha\!(\alpha,\beta,\Theta_2,\Theta_3,\Theta_4)\, V_A(b) \,,\\
\tau_\beta & = \frac{1}{4}\big(b_a,_\beta\!(\beta,\Theta_1) \,V_A(a)+ 2 b_b,_\beta\!(\alpha,\beta,\Theta_2,\Theta_3,\Theta_4) \,V_A(b)\big)\,,
\\
\Tc_1 & = \frac{1}{2}\,b_a,_{\Theta_1}\!(\beta,\Theta_1) \,V_A(a)\,,\\
\Tc_2 & =  \frac{1}{2}\,b_b,_{\Theta_2}\!(\alpha,\beta,\Theta_2,\Theta_3,\Theta_4) \,V_A(b)\,,\\
\Tc_3 & = b_b,_{\Theta_3}\!(\alpha,\beta,\Theta_2,\Theta_3,\Theta_4)\, V_A(b)\,,\\
\Tc_4 & = b_b,_{\Theta_4}\!(\alpha,\beta,\Theta_2,\Theta_3,\Theta_4) \,V_A(b)\,.
\end{aligned}
\end{equation}

\section{Analytical \textit{\lowercase {vs}} LAMMPS results}
The most direct outcomes of our solution are  natural geometry and  energy, which can be used to check the correctness of whatever MD code. The results obtained by solving equations \eqref{finali} with Brenner 2nd generation potential are in good agreement with First Principles, Density Functional Theory (DFT) and Diffusion Monte Carlo (DMC) simulations, as Tables  \ref{tab:r:comparisonen} and \ref{tab:r:comparison} show.

As an application of the possibility of exploiting \eqref{finali} as a benchmark, we present in Table \ref{tab:r} the radii of a number of CNTs, showing that standard LAMMPS code underestimates them. In Table \ref{tab:cohesive} the values of the cohesive energy from our solution and those obtained with the use of LAMMPS code when adopting the 2nd-generation Brenner potential are presented; it can be seen that there is a remarkable difference: the cohesive energy  is highly overestimated and our benchmark makes evident that the code in its current implementation definitely produces   results which are offset from those of the benchmark. The origin of the discrepancies can be found only by a close inspection of LAMMPS source code. In fact, although in Brenner \textit{ et al.}\cite{Brenner_2002} it is indicated that the values of the function $P_{IJ}$ should be taken null for solid-state carbon, the code assigns the value 0.027603.  This latter value is actually dictated in Table VIII of  Stuart \textit{et al.}\cite{Stuart2000} for AIREBO potentials,  due to the additional terms included in this potential. Whenever a LAMMPS user wants to adopt REBO potentials, he needs to change the hard-wired number for the variable PCCf[2][0] in ``pair\_airebo.cpp''; unfortunately,  the LAMMPS manual does not provide any information on this issue, and most studies based on LAMMPS REBO calculations are likely to have underestimation or overestimation of mechanical and geometrical properties presented in our Tables.  An example of the use of LAMMPS with 2nd generation Brenner potential is Zhang \textit{ et al.}\cite{Zhang2014}.
When the  value assigned in Brenner \textit{ et al.}\cite{Brenner_2002} is implemented, the LAMMPS code produces  the same results as the benchmark solution, letting alone a tiny difference due to numerical effects, as the third column of Tables \ref{tab:r} and \ref{tab:cohesive} undeniably makes evident. 

Starting from the geometry and the energy gathered by means of \eqref{finali}, it is possible to obtain secondary quantities.	In Table \ref{tab:young}  and \ref{tab:poisson} the Young moduli and the Poisson coefficients are reported: the standard LAMMPS code overestimates the former and underestimates the latter. Our results are in very good agreement with the literature (see e.g. Agrawal \textit{ et al.}\cite{Agrawal_2006}). The differences between our benchmark and the LAMMPS code with modified parameters are ascribable to numerical effects, more accentuated  because Young modulus and Poisson coefficients are  quantities not directly evaluated, but rather derived, and an  increment of numerical error is foreseeable.

\section*{Acknowledgments}
AF  acknowledges the Italian INdAM-GNFM (Istituto Nazionale di Alta Matematica -- Gruppo Nazionale di Fisica Matematica), through ``Progetto Giovani 2014 --  Mathematical models for complex nano- and bio-materials''.  NP is supported by the European Research
Council (ERC StG Ideas 2011 BIHSNAM n. 279985 on ``Bio-Inspired hierarchical super-nanomaterials'', ERC PoC 2013-1 REPLICA2 n. 619448 on ``Large-area replication of biological antiadhesive nanosurfaces'', ERC PoC 2013-2 KNOTOUGH n. 632277 on `Super-tough knotted fibres''), by the European Commission under the Graphene Flagship (WP10 ``Nanocomposites'', n. 604391) and by the Provincia Autonoma di Trento (``Graphene Nanocomposites'', n. S116/2012-242637 and reg. delib. n. 2266). 



{\tiny
	\begin{table}[h!]
		\begin{center}
			\caption{Cohesive eneregy (eV/atom)}
			\label{tab:r:comparisonen}
			\vskip 10pt
			\begin{tabular}{ccccc}
				our  & El-Barbery & Shin   \\
				benchmark  & et al. \cite{Elbarbary_2003} & et al. 2014 \cite{Shin_2014} \\
				&  (First Principles) & (DMC)     \\
				\hline -7.3951 & -7.4  &  -7.464  \\
				\hline
			\end{tabular}
		\end{center}
	\end{table}

	\begin{table}[h!]
		\begin{center}
			\caption{Radii (nm) of small CNTs, comparison with literature}
			\label{tab:r:comparison}
			\begin{tabular}{ccccccccc}
				\ & \ &  \ & \ &  \ & \ & \ \\
				$(n,m)$ & our  &  Mach\'on & Cabria &  Popov & Budyka \\
				\ & benchmark &  et al. 2002 \cite{Machon2002} & et al. 2003 \cite{Cabria2003} &   2004 \cite{Popov2004} & et al. 2005\cite{Budyka_2005}   \\
				\ &  & (DFT) & (DFT) & (TB)  & (DFT)  \\[3pt]
				\hline
				(3,3) & 0.211 &  0.210 & 0.212 & 0.212 & -      \\
				(4,4) & 0.277 &  -     & -     & -     & 0.277     \\
				(5,0) & 0.208 &  0.204 & 0.206 & 0.205 & -        \\
				\hline
			\end{tabular}
		\end{center}
	\end{table}

	\begin{table}[h]
		\begin{center}
			\caption{Radii}
			\label{tab:r}
			\vskip 3pt
			\begin{tabular}{cccc}
				\ & Our   & LAMMPS  & LAMMPS  \\[3pt]
				\ & benchmark &  (standard) & (modified)   \\[3pt]
				$(n,m)$ & (nm) & (nm) & (nm)    \\[3pt]
				\hline
				(3,3)   &  0.2111 & 0.2079 &  0.2110\\
				(4,4)   & 0.2767 &  0.2723 & 0.2766\\
				(5,5)   & 0.3431 & 0.3371 & 0.3404\\
				(6,6)   & 0.4101 & 0.4035 & 0.4100\\
				(7,7)   & 0.4773 & 0.4697 & 0.4773\\
				(8,8)   & 0.5447 & 0.5361 & 0.5447\\
				(10,10) & 0.6798 & 0.6690 & 0.6798 \\
				(12,12) & 0.8151 & 0.8022 & 0.8151\\
				(18,18) & 1.2216 & 1.2020 & 1.2215\\
				(25,25) & 1.6961 & 1.6689  & 1.6960   \\
				\hline
				(5,0) & 0.2078 &  0.2046 & 0.2076 \\
				(6,0) & 0.2447 & 0.2409 & 0.2446\\
				(7,0) & 0.2823 & 0.2778 & 0.2821 \\
				(8,0) & 0.3202 & 0.3151 & 0.3201\\
				(9,0) & 0.3584 & 0.3527 & 0.3583\\
				(10,0) & 0.3969 & 0.3905 & 0.3967\\
				(12,0) & 0.4741 & 0.4665 & 0.4739\\
				(15,0) & 0.5905 & 0.5810 & 0.5904\\
				(20,0) & 0.7853 & 0.7274 & 0.7852\\
				(30,0) & 1.1760 & 1.1572 & 1.1759\\
			\end{tabular}
		\end{center}
	\end{table}
	
	\begin{table}[h]
		\begin{center}
			\caption{Cohesive energy}
			\label{tab:cohesive}
			\vskip 3pt
			\begin{tabular}{cccc}
				\ & Our   & LAMMPS  & LAMMPS  \\[3pt]
				\ & benchmark &  (standard) & (modified)   \\[3pt]
				$(n,m)$ & (eV/atom) & (eV/atom) & (eV/atom)    \\[3pt]
				\hline
				(3,3)   & -7.0137 & -7.3838 & -7.0137  \\
				(4,4)   & -7.1695 &  -7.5569 & -7.1695\\
				(5,5)   & -7.2463 & -7.6422  & -7.2462 \\
				(6,6)   & -7.2898 & -7.6905 & -7.2896\\
				(7,7)   & -7.3167 & -7.7204 & -7.3166\\
				(8,8)   & -7.3346 & -7.7403 & -7.3345\\
				(10,10) & -7.3560 & -7.7640 & -7.3558\\
				(12,12) & -7.3678 & -7.7771& -7.3676\\
				(18,18) & -7.3829 & -7.7038 & -7.3827 \\
				(25,25) & -7.3887 & -7.8003 & -7.3886    \\
				\hline
				(5,0) & -6.9758 &  -7.3417 & -6.9759\\
				(6,0) & -7.0969 & -7.4763 & -7.0969 \\
				(7,0) & -7.1715 & -7.5593 & -7.1715\\
				(8,0) & -7.2212 & -7.6144 & -7.2212\\
				(9,0) & -7.2560 & -7.6531 & -7.2560\\
				(10,0) & -7.2814 & -7.6812 & -7.2813\\
				(12,0) & -7.3151 & -7.7186 & -7.3149\\
				(15,0) & -7.3432 & -7.7499 & -7.3431\\
				(20,0) & -7.3656 & -7.7747& -7.3655\\
				(30,0) & -7.3819 & -7.7927 & -7.3818\\
				\hline
				graphene & -7.3951  & -7.8074 & -7.3950
			\end{tabular}
		\end{center}
	\end{table}

	\begin{table}[h]
		\begin{center}
			\caption{Young modulus}
			\label{tab:young}
			\vskip 3pt
			\begin{tabular}{cccc}
				\ & Our   & LAMMPS & LAMMPS   \\[3pt]
				\ & benchmark &  (standard)  & (modified)   \\[3pt]
				$(n,m)$ & (GPa) & (GPa) & (GPa)  \\[3pt]			\hline
				(3,3) &  893.9167  & 987.0102 & 885.0631 \\
				(4,4) & 851.4536 & 944.4810 & 840.1351\\
				(5,5) &  804.6053 & 901.869 & 799.7630\\
				(6,6) & 800.9298 & 891.6427 & 789.8102\\
				(7,7) & 793.7379 & 881.2895 & 778.2888\\
				(8,8) &  784.8784 & 872.9931& 767.9165\\
				(10,10) & 768.6379  & 856.9461 & 756.7625\\
				(12,12) & 756.3044  & 846.2911 & 746.5046\\
				(18,18) & 735.9332   & 831.2163 & 732.5252\\
				(25,25)& 726.2650  & 823.3865 & 726.4968\\
				\hline
				(5,0) &  948.0854 & 1046.3569 & 943.9120\\
				(6,0) & 973.0048 & 1075.4912 & 968.5679\\
				(7,0) & 976.5265 & 1082.0265 & 971.5102\\
				(8,0) & 969.7954  & 1076.6910 & 965.8188\\
				(9,0) & 958.1824  & 1066.6410 & 954.9396\\
				(10,0) & 944.5151 & 1053.1830 & 941.3135\\
				(12,0) & 916.0510 & 1025.8253 & 915.1339\\
				(15,0) & 877.5820  & 986.5745 & 877.9641\\
				(20,0) &  830.0108 & 940.5973 & 835.7993\\
				(30,0) & 779.7321 & 890.5147 & 789.0504\\
			\end{tabular}
		\end{center}
	\end{table}

	\begin{table}[h]
		\begin{center}
			\caption{Poisson coefficient}
			\label{tab:poisson}
			\vskip 3pt
			\begin{tabular}{cccc}
				\ & Our   & LAMMPS & LAMMPS   \\[3pt]
				$(n,m)$  & benchmark &  (standard)  & (modified)   \\[3pt]
				\hline
				(3,3) & 0.1450  & 0.1237 & 0.1563  \\
				(4,4) & 0.2311  & 0.2078 & 0.2388 \\
				(5,5) & 0.2924 &  0.25782  & 0.2963         \\
				(6,6) & 0.3061 & 0.27936  & 0.3115\\
				(7,7) & 0.3181 &  0.2937 & 0.3318	\\
				(8,8) & 0.3292  & 0.3020 & 0.3458\\
				(10,10) & 0.3466 & 0.3194 & 0.3526\\
				(12,12) &  0.3588 &  0.3306 & 0.3626\\
				(18,18) & 0.3779  & 0.3426 & 0.3740\\
				(25,25)& 0.3867  & 0.3507 & 0.3790\\
				\hline
				(5,0) & 0.0655 & 0.0362 & 0.0661  \\
				(6,0) & 0.0867 & 0.0600 & 0.0840\\
				(7,0) & 0.1100 & 0.0800 & 0.1105\\
				(8,0) & 0.1328  & 0.1045 & 0.1306\\
				(9,0) &  0.1544 & 0.1234 & 0.1511\\
				(10,0) & 0.1742 & 0.1424 & 0.1737 \\
				(12,0) & 0.1923 & 0.1725 & 0.2032\\
				(15,0) &  0.2493 & 0.2138 & 0.2446\\
				(20,0) & 0.2950  & 0.2564 & 0.2835\\
				(30,0) & 0.3406 & 0.2991 & 0.3261
			\end{tabular}
		\end{center}
	\end{table}
	
}

\end{document}